# Binding of Carminomycin to synthetic polyribonucleotides poly(A) and poly(U): absorption and polarized fluorescence study

**N.N. Zhigalova [1], V.N. Zozulya [2], O.A. Ryazanova [2]**

*[1] Department of Molecular and Medical Biophysics, School of Radiophysics, Biomedical Electronics and Computer Systems, V.N. Karazin Kharkiv National University, 4 Svobody sq., 61022 Kharkiv, Ukraine. E-mail: zhigalova@karazin.ua*

*[2] Department of Molecular Biophysics, B. Verkin Institute for Low Temperature Physics and Engineering, National Academy of Sciences of Ukraine, 47 Nauky ave, 61103 Kharkiv, Ukraine. E-mail: ryazanova@ilt.kharkov.ua*

Binding of anthracycline antibiotic carminomycin (CM) to synthetic polyribonucleotides poly(A) and poly(U) was studied in solution of low ionic strength in a wide phosphate-to-dye (*P/D*) range using absorption and polarized fluorescence spectroscopy. Two different modes of CM binding to the *ss*-polynucleotides have been identified. The first of them dominating at low phosphate-to-dye (*P/D*) ratios is self-assembly of the heterocyclic dye on the polymer surface driven by to cooperative electrostatic binding of amino group of CM sugar moiety to negatively charged polynucleotide phosphate groups with the chromophores self-stacking. At high *P/D* values, the stacking-associates disintegrate and monomeric binding of ligand to nucleic bases become prevalent. Thermodynamic parameters of binding were estimated for the both cases.

## INTRODUCTION

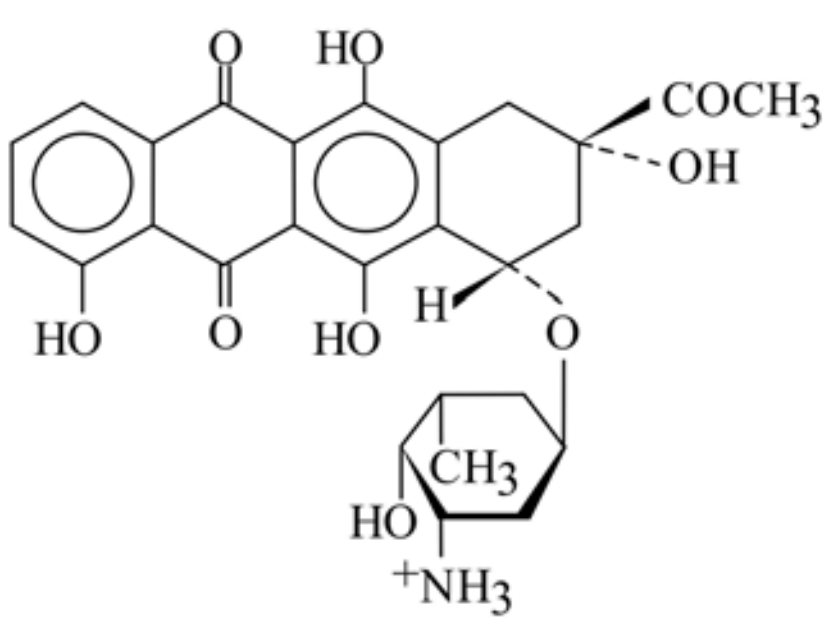

**Fig. 1.** Molecular structure of carminomycin (CM).

It is well known that binding of the heterocyclic biologically active compounds with a planar molecular geometry (dyes, antibiotics, carcinogens, photosensitizers and some others) to nucleic acids (NA) in most cases realizes via intercalation of their chromophores between nucleic bases of double-stranded DNA that disrupt the transmission of hereditary information and leads to the cell death. This is also one mechanism of action of antitumor anthracycline antibiotics.

However, in addition to intercalation, molecules of the cationic dyes can also bind electrostatically to polyanionic phosphate backbone of NA with or without self-stacking. In most cases, the focus was on the strong intercalation binding type, whereas weak external binding was often neglected despite for single-stranded RNA and DNA molecules with readily available phosphate groups and unpaired nucleic bases this binding type can be predominant.

Antracycline antibiotic carminomycin known as selective inhibitor of nucleic acids synthesis in cells of microorganisms and malignant tumors [1] forms complexes with DNA *in vitro* and considerably increases the melting temperature of DNA. Also this antibiotic inhibits the template activity of DNA in the system of DNA-dependent RNA polymerase, as well as repair in bacterial cells injured by radiation and alkylating agents [2,3]. Data on the physical and chemical properties of CM as well as on mechanisms of its action are given in the articles [4,5,6,7,8,9]. Earlier the binding of CM to native double-helical and denatured DNA from chicken erythrocites [10], calf thymus DNA and covalently closed circular PM-2 DNA [11, 12] have been studied. However, the complexation of this anthracycline antibiotic with single-stranded polynucleotides is insufficiently investigated.

In this work binding of carminomycin (Fig. 1) to synthetic poly(A) and poly(U) homopolyribonucleotides has been studied using methods of molecular spectroscopy, namely, absorption and polarized fluorescence spectroscopy techniques. The last one is known as a very efficient method of studying the intermolecular interactions of aromatic compounds, because the ligand fluorescence intensity is very sensitive to changes in its microenvironment, whereas the fluorescence polarization degree reflects the mobility of the chromophore, which always decreases upon binding [13,14,15]. The conformational properties of the polynucleotides studied allow us to model different structural matrices for intercalator binding. So, depending on the solution pH poly(A) can take shape of ordered single or double-stranded helices, whereas poly(U) in all pH range represents single chain with disordered bases. The obtained experimental data allowed us to suggest models for the binding of CM to single- and double-stranded matrices and to estimate quantitatively the ratio between inputs of two types of CM binding to polynucleotides, namely, with nucleobases and phosphate groups.

## MATERIALS AND METHODS

Carminomycin hydrochlorid (CM, Fig. 1) from FSBI Gause Institute of New Antibiotics (Moscow, Russia) was used as received. Other reagents were purchased from Sigma (Germany). The concentrations of the dye and polyribonucleotides were determined spectrophotometrically using the molar extinction coefficients of $\varepsilon_{492} = 14{,}500$ $M^{-1}cm^{-1}$ for CM [16], $\varepsilon_{257} = 10{,}100$ $M^{-1}cm^{-1}$ for poly(A) and $\varepsilon_{260} = 9{,}600$ $M^{-1}cm^{-1}$ for poly(U) [17].

For all experiments 1 mM sodium cacodylate buffer containing 0.5 mM $Na_2EDTA$ was prepared using fresh deionized distilled water. Experiments with single-stranded polynucleotide were carried out in solution of pH 7, whereas complexes with double-stranded poly(A) was studied in the buffer of pH 4.9. Buffers were ultra filtered through nitrocellulose filters (Millipore-Q system, USA) with a pore diameter of 0.22 microns.

Samples for spectral measurements were prepared by titration of carminomycin sample ($C_{dye} = 1.7 \cdot 10^{-5}$ M) with increasing amount of concentrated polymer solution containing the same CM content in predetermined volume proportions that allowed to get desired molar phosphate-to-dye ratios, *P/D*. Measurements were carried out in quartz cuvettes at ambient temperature of 22–24 °C.

Electronic absorption spectra were measured on a SPECORD UV-VIS spectrophotometer (VEB Carl Zeiss, Jena).

Steady-state fluorescence measurements were carried out by the method of photon counting with a laboratory spectrofluorimeter based on double monochromator DFS-12 (LOMO, Russia) [10]. Fluorescence excitation was performed by linearly-polarized beam of He-Cd laser LPM-11 ($\lambda_{exc} = 441.6$ nm), which power was stabilized during the experiment using hand-made set-up described in [18]. The fluorescence intensity was registered at the maximum of CM fluorescence band ($\lambda_{obs} = 584$ nm) at the right angle to the incident beam. Ahrens prisms were used to polarize linearly the exciting beam as well as to analyze the fluorescence polarization. The spectrofluorimeter was equipped with a quartz depolarizing optical wedge to exclude the monochromator polarization-dependent response. When measuring the fluorescence intensity, the pulses from photomultiplier tube were accumulated during 10 s for each data point and measurements were repeated five times, at that the measurements error was about 0.5%. Also the correction was made to the absorption of the laser beam in the solution layer between the front wall and the center of the cuvette. Fluorescence spectra were corrected on the spectral sensitivity of the spectrofluorimeter. Experimental set-up and the measurement procedure were described earlier [10].

The total fluorescence intensity, *I*, its polarization degree, *p*, and anisotropy, *μ*, were calculated using formulas [13]:

$$I = I_{II} + 2I_{\perp} \tag{1}$$

$$p = \frac{I_{II} - I_{\perp}}{I_{II} + I_{\perp}} \tag{2}$$

$$\mu = \frac{I_{II} - I_{\perp}}{I} = \frac{2p}{3-p} \tag{3}$$

where $I_{II}$ and $I_{\perp}$ - are measured intensities of the emitted light, which are polarized parallel and perpendicular to the polarization direction of the exciting light beam, respectively.

.

## RESULTS AND DISCUSSION

Absorption spectrum of free CM in aqueous solution represents wide asymmetric band with maximum at 493 nm, whereas maximum of fluorescence band with slightly splitted top lies at 584 nm [16].

Figs. 2 and 3 represent relative absorbance, $A/A_0$, relative fluorescence intensity, $I/I_0$, and fluorescence polarization degree, *p*, *vs P/D* upon titration of CM with poly(A) and poly(U), where $A_0$ , $I_0$ correspond to properties of the dye in a free state (unbound dye).

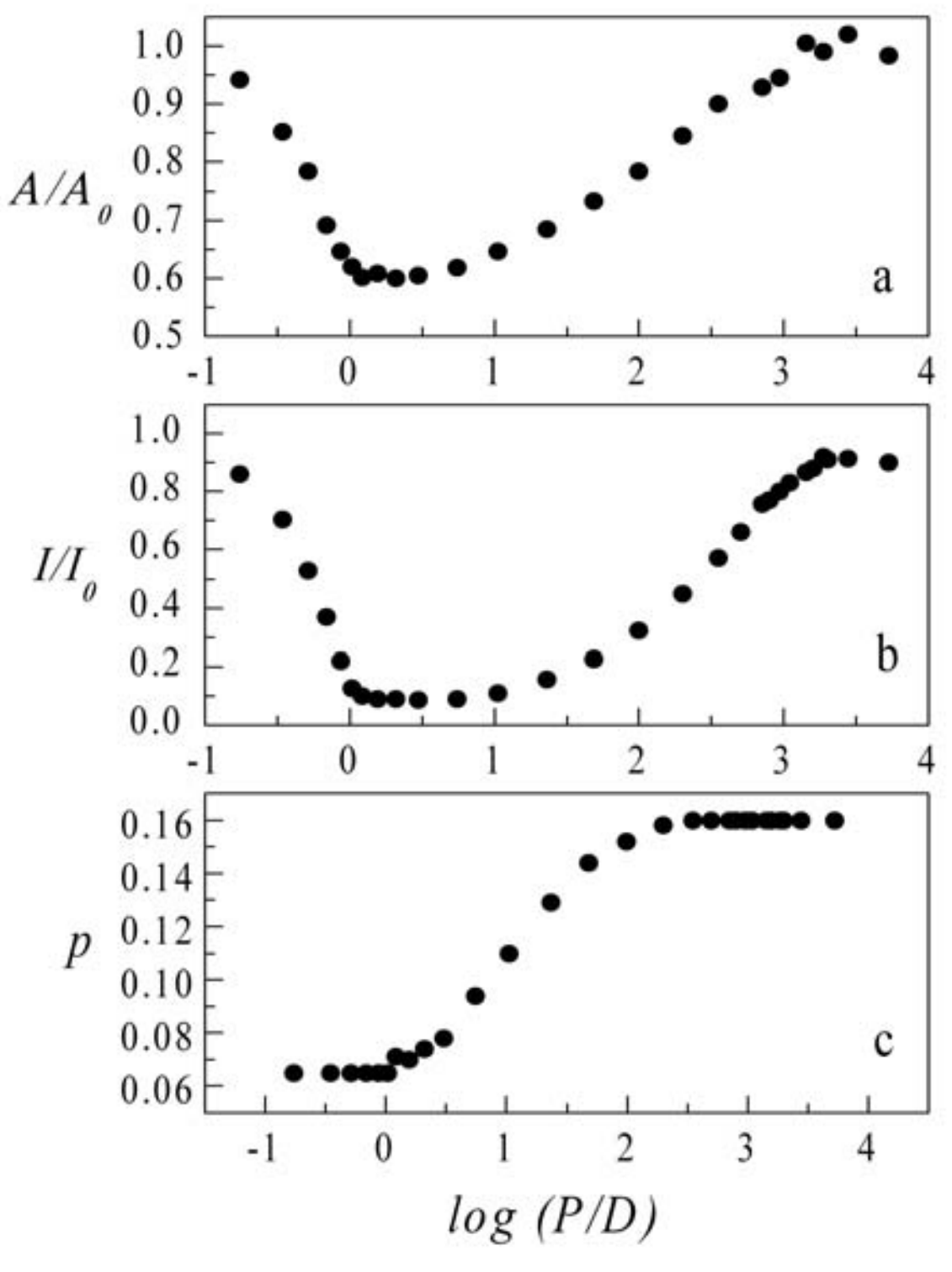

**Fig. 2**. Changes in absorbance (a), fluorescence intensity (b) and fluorescence polarization degree (c) of CM upon binding to poly(U) in neutral solution (pH7).

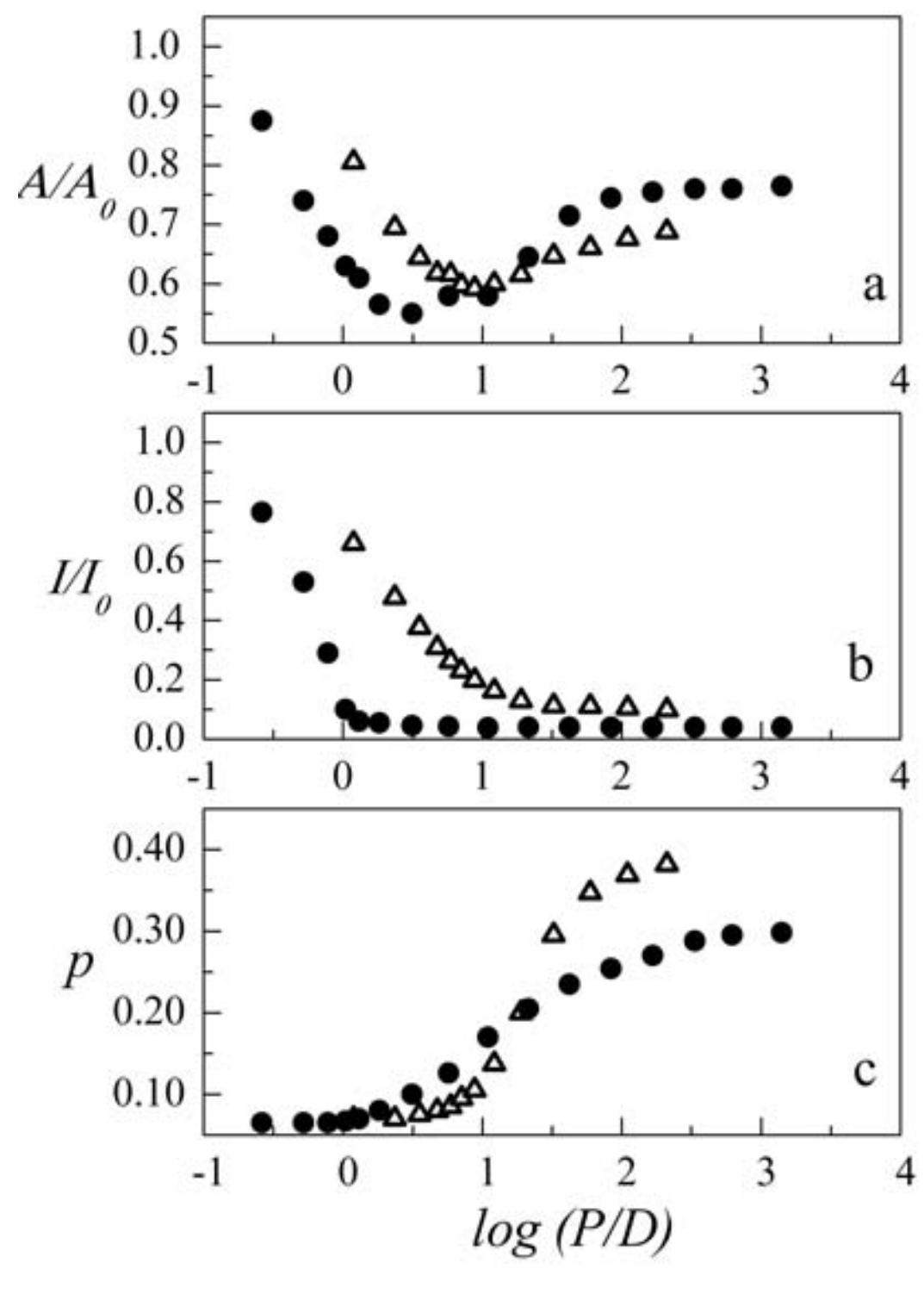

**Fig. 3.** Changes in absorbance (a), fluorescence intensity (b) and fluorescence polarization degree (c) of CM upon binding to *ss*-poly(A) in neutral solution pH7 (●) and *ds*-poly(A) in acidic solution pH 4.9 (Δ).

Biphasic shape of fluorimetric titration curves evidences formation of two different types of complexes between CM and the single-stranded polynucleotides. From strong fluorescence quenching characterizing by linear dependence of emission intensity *vs* *P/D* (Figs. 2b, 3b) and practically constant value of fluorescence polarization degree (Figs. 2c, 3c) observed in the range of low *P/D* ratios (*P/D* < 2) it was concluded that fluorescence is emitted only by unbound dye molecules, whereas for bound dye it is completely quenched. It is seen that binding is accompanied by substantial decrease in the sample absorbance (Figs. 2a, 3a, Table 1). The similar spectroscopic changes were observed earlier [16] upon carminomycin binding to polyanionic chain of inorganic polyphosphate, poly(P), representing a good model of nucleic acids backbone, where they were explained by cooperative binding of CM molecules to the polymer surface with self-stacking. Therefore, it can be concluded that interaction of CM molecules to the polynucleotides studied at low *P/D* is determined mainly by Coulomb attraction between the dye aminosugar and the polynucleotide phosphates as well as by van der Waals forces determining self-stacking of the dye chromophores.

At the same time, highly polarized emission of CM observed at high *P/D* ratios evidences strong interaction between the chromophores and polynucleotide bases as a result of gradual disintegration of externally bound CM stacks into monomers whose amino groups probably remain electrostatically bound to the polymer phosphates. At very high *P/D* ratios spectroscopic properties of CM can be attributed to its complexes with the nucleobases, that is confirmed by the high value of fluorescence polarization degree as results of deceleration of rotational motion for the bound molecules, as well as by absorption changes (3 nm red shift of absorption band maximum upon binding to poly(U) and 10 nm one in the case of poly(A)). It can be noted that emission intensity and absorbance of CM bound to uracil bases at very high *P/D* are the practically the same as for the free dye, whereas for similar complexes with adenines substantial absorption hypochromism and strong fluorescence quenching were observed that evidences the stacking between dye chromophores and adenine bases. The mechanism of quenching of CM emission by adenine was explained earlier in [19].

For comparison, an interaction of CM with poly(A) was studied also in the solution of pH 4.9, where this polynucleotide takes conformation of a double helix. As it is seen from Fig. 3, under such experimental condition adenine also induces strong quenching of CM emission. Substantial increase in fluorescence polarization degree, hypochromism and 13 nm red shift of CM absorption band evidences intercalation binding mechanism. The dependence of the CM fluorescence intensity on *P/D* in its complex with *ds*-poly(A) is the same as that observed earlier for native DNA [10].

Spectroscopic properties of CM bound to nucleic bases (at high *P/D*) were identified as limit values of their dependencies on *D/P*. They are summarized in the Table 1 along with data characterizing outside binding of the dye to the polynucleotides exterior at low *P/D*.

**Table 1.** Spectroscopic properties of CM in a free state and bound to the biopolymers in solution of low ionic strength (μ = 0.002).

| Sample | Low *P/D* ratios (*P/D* = 2-10) | | | High *P/D* ratios | | |
|---|---|---|---|---|---|---|
| | $A/A_0$ minimal values | $I/I_0$ | $p$ | $A/A_0$ | $I/I_0$ | $p$ |
| CM | 1.00 | 1.00 | 0.065 | 1.00 | 1.00 | 0.065 |
| CM + poly(U) | 0.60 [a] | 0.086 [a] | 0.073 [a] | 1.00 | 0.90 | 0.160 |
| CM + *ss*-poly(A) | 0.55 [b] | 0.042 [b] | 0.099 [b] | 0.76 | 0.04 | 0.300 |
| CM + *ds*-poly(A) | 0.59 [c] | 0.178 [c] | 0.117 [c] | 0.69 | 0.10 | 0.380 |
| CM + poly(P) [16] | 0.57 [a] | 0.02 [a] | 0.062 [a] | 0.82 | 0.70 | 0.085 |
| CM + nDNA [d] [10] | – | – | – | 0.72 | 0.004 | 0.350 |
| CM + dDNA [e] [10] | – | – | – | – | 0.012 | 0.210 |

[a] values at *P/D* = 2 (minimum of absorption titration curve)
[b] values at *P/D* = 3 (minimum of absorption titration curve)
[c] at *P/D* = 10 (minimum of absorption titration curve)
[d] double-stranded native DNA from chicken erythrocytes
[e] denatured (single-stranded) DNA

The value of fluorescence polarization degree is known as an important parameter characterizing binding process which is strongly affected by rotational mobility of the polynucleotide bases coupled with dye. As it is seen from Table 1, for the more labile disordered poly(U) chain $p$ takes the lowest value, 0.16, for ordered single-stranded poly(A) taking conformation of A-type helix with partially stacked adenine bases $p$ = 0.30, whereas for rigid double-stranded poly(A) it is highest, 0.38. From aforesaid experimental data an intercalation of CM chromophores between adenine bases of *ds*-poly(A) and their partial intercalation (pseudointercalation) between nucleobases of *ss*-poly(A) at high *P/D* ratios can be suggested.

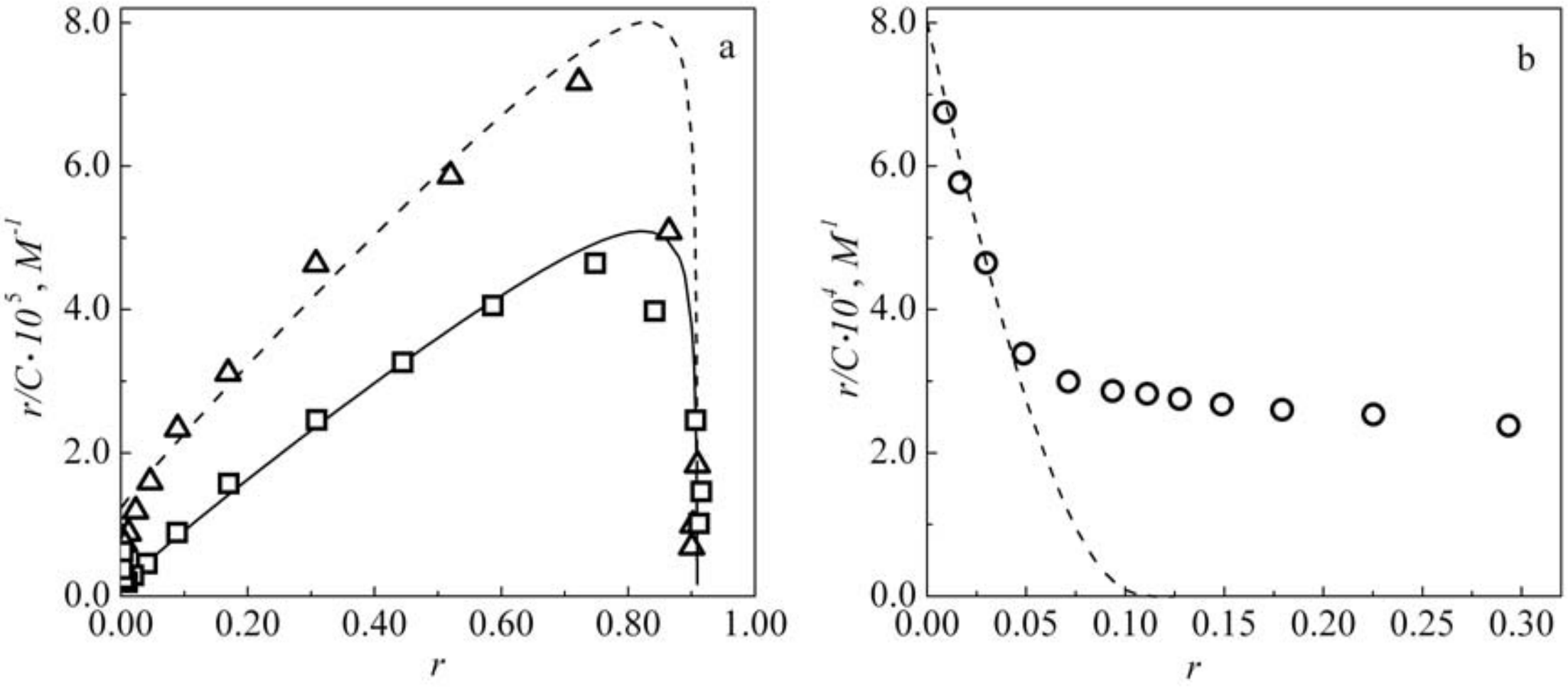

**Fig. 4.** Scatchard plot for carminomycin binding to (**a**) single-stranded poly(U) (□) and poly(A) (Δ) in solution pH 7; (**b**) double-stranded poly(A) (○) in solution pH 4.9. Solid and dashed lines represent approximation of experimental data with McGhee and von Hippel equations (7) and (8) with binding parameters listen in the text below.

Quantitative characteristics of complex formation were obtained from analysis of binding isotherms plotted in Scatchard coordinates (Fig. 4), where $r = \gamma_b \cdot D / P$ is the number of bound CM molecules per one nucleotide, and $C = (1 - \gamma_b)D = \gamma_0 D$ is the molar concentration of free CM in solution. Fraction of bound carminomycin, $\gamma_b$, was calculated using Ellerton and Isenberg equation [20]:

$$\gamma_b = I / I_0 \cdot [(\mu - \mu_b)/(\mu_0 - \mu_b)] \qquad (4),$$

where $\mu_0$ and $\mu_b$ are anisotropy values for free and bound dye molecules.

The convex shape of Sketchard curves observed in the case of single-stranded polynucleotides (Fig. 4a) indicates strong cooperative character of the CM electrostatic binding to the polyribonucleotides with self-stacking of the dye chromophores. The apparent equilibrium binding constant and size of binding sites can be determined using the Schwarz method [21] where the changes in the fraction of free ligand, $\gamma_0$, were analyzed at low *P/D* ratios. Since the emission of CM molecules electrostatically bound to polynucleotide exterior with self-stacking is supposed to be fully quenched, the initial section of the fluorescence titration curves (Fig. 5) represents just a dependence $I/I_0 = \gamma_0$. So, linear parts of these curves correspond to the full binding saturation. The point of intersection of the extrapolation line with the *P/D* axis determines the number of the polynucleotide phosphates per binding sites, *n*, which for poly(U) and poly(A) is the same, *n* = 1.1. The values of apparent binding constants for CM external cooperative binding were determined according formula:

$$K = 1 / D\gamma_0^*, \qquad (5)$$

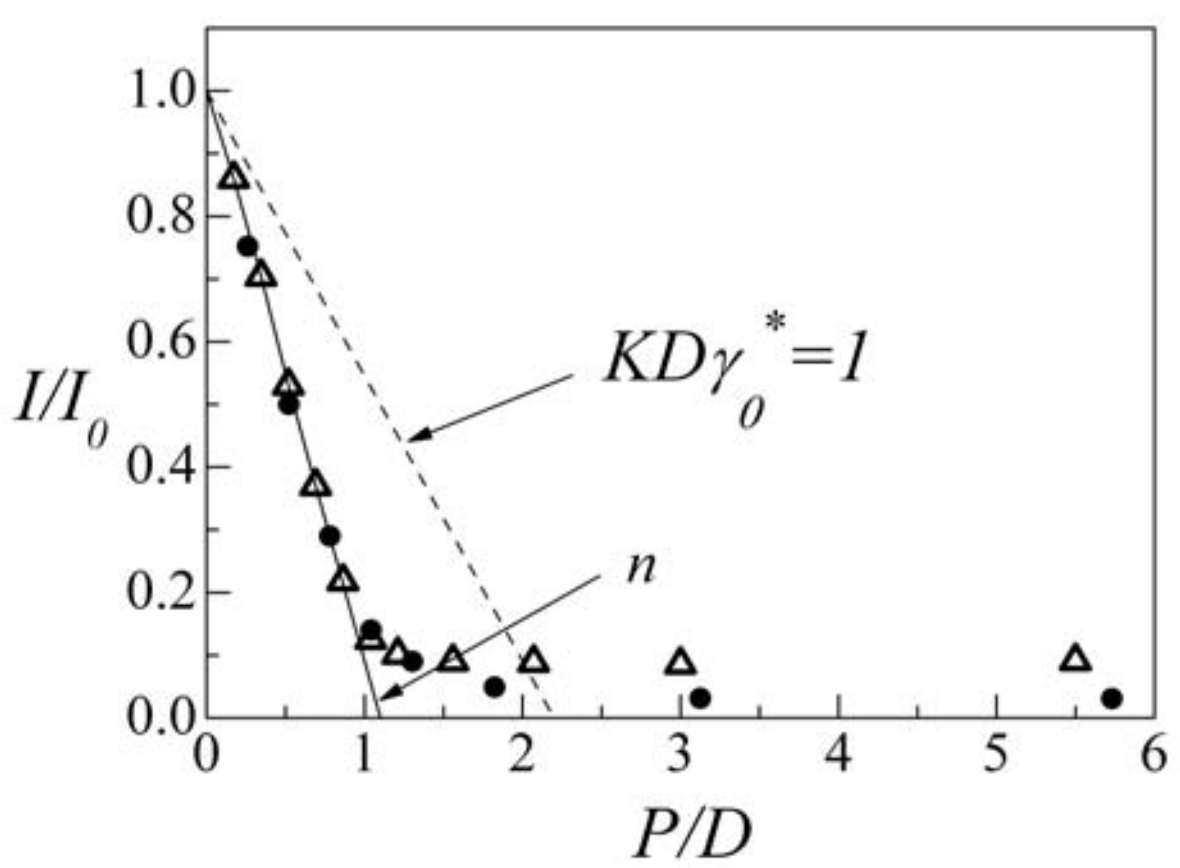

**Fig. 5.** Fluorimetric titration curves plotted as dependence of relative fluorescence intensity *vs P/D* for carminomycin complexes with *ss*-poly(A) (●) and poly(U) (Δ) in solution pH7, Total dye concentration is constant, $C_{CM} = 1.7 \cdot 10^{-5}$ M, $\lambda_{exc} = 441.6$ nm; $\lambda_{obs} = 584$ nm. Solid and dashed lines correspond to determination of binding parameters by Schwarz method.

where $\gamma_0^*$ is the fraction of the ligand in a free state when the polyanionic lattice is half occupied. The $\gamma_0^*$ values were determined as the intercection points of fluorimetric titration curves plotted as $I/I_0$ versus $P/D$ curves with the straight line drawn at half the negative slope of the stoichiometric part of these curves (dotted line in Fig. 5). They are equal to $\gamma_0^* = 0.09$ for poly(U) and 0.05 for poly(A). By substitution of these values and CM concentration into the formula (5), we have calculated the values of the cooperative binding constants, $K$, which are equal to $6.5 \cdot 10^5$ M$^{-1}$ for poly(U) and $1.16 \cdot 10^6$ M$^{-1}$ for poly(A). At the same time, the constant of cooperative binding can be presented as:

$$K = K_1 \cdot \omega \qquad (6),$$

where $K_1$ is apparent equilibrium binding constant for monomeric binding, $\omega$ is the mean cooperativity parameter.

Experimental data plotted in Scatchard coordinates were fitted with McGhee and von Hippel equation [22], which takes into account two types of binding to the one lattice: (i) cooperative self-stacking of CM on the surface of the polymers and (ii) non-cooperative binding of the dye to nucleic bases:

$$\frac{r}{C} = (1 - nr)\left[K_2 + K_1\left[\frac{(2\omega - 1)(1 - nr) + r - R}{2(\omega - 1)(1 - nr)}\right]^{(n-1)}\left[\frac{1 - (n+1)r + R}{2(1 - nr)}\right]^2\right]$$

$$R = \left[\left[1 - (n+1)r\right]^2 + 4\omega r(1 - nr)\right]^{1/2} \qquad (7)$$

Here, an apparent equilibrium constant $K_2$ characterizes just binding of CM to nucleic bases. For both types of complexes the same size of binding sites, $n$=1.1, was taken, it was defined as the number of nucleotides per one adsorbed CM molecule, The intersection points of experimental Skatchard plots with the $r/C$ axis at $r = 0$ (Fig. 4a) give the sum of the monomer binding constants, $K_1$+$K_2$, equal to $10^4$ $M^{-1}$ and $1.2 \cdot 10^5$ $M^{-1}$ for CM complex with poly(U) and poly(A), correspondingly. $K_1$ was supposed to be equal to the value of the monomeric electrostatic binding constant of CM to inorganic polyphosphate, $K_1 = 1.8 \cdot 10^3$ $M^{-1}$ [16]. So, $K_2$ can be easily determined from the sum of constants taking the values of $8.2 \cdot 10^3$ $M^{-1}$ for poly(U) and $1.18 \cdot 10^5$ $M^{-1}$ for poly(A). Unknown parameter $\omega$ was determined from better fit being equal to 360 for CM complex with poly(U) and 600 in the case of poly(A). Hence, the equilibrium constants of CM cooperative binding to poly(U) and poly(A) calculated from equation (6) are equal to $6.5 \cdot 10^5$ $M^{-1}$ and $1.08 \cdot 10^6$ $M^{-1}$ correspondingly, those are in a good agreement with the values obtained by·Schwartz method.

It is seen, that he parameters of two types of CM binding to the polyribonucleotides studied are different. The magnitude of cooperativity parameter is greater in the case of poly(A), since linear density of its negative charges is higher. Also, the apparent binding constant of CM with adenine bases of poly(A) is much greater than that for uracil bases of poly(U). It can be explained by the partially ordered secondary structure of poly(A) helix, where the adenine bases are partially stacked, whereas the nucleobases in poly(U) are completely disordered.

For CM binding to double-stranded poly(A) the Scatchard plot (Fig. 4b) also evidences existence of two binding types, where intercalative binding is predominant. The steep part of binding isotherm corresponding to CM intercalation can be well described by McGhee and von Hippel equation for non-cooperative binding:

$$\frac{r}{C} = K \frac{(1 - nr)^n}{[1 - (n-1)r]^{n-1}} \qquad (8),$$

using parameters of $n = 8$, $K = 8 \cdot 10^4$ $M^{-1}$. The second part of the isotherm corresponds to a weak electrostatic binding to the external polynucleotide surface. The similar behavior was observed earlier upon CM binding to native DNA [10].

## CONCLUSIONS

Cooperative electrostatic binding plays an important role in the interaction of carminomycin with single-stranded nucleic acids. Binding of CM to nucleic bases of single-stranded polynucleotides can be realized only with simultaneous electrostatic binding of the dye aminosugar to the biopolymer phosphates. Whereas upon interaction of CM with *ds*-poly(A) the intercalative binding is predominant, and input of electrostatic binding mode is less significant.